\documentclass{nature}

\usepackage{subcaption}
\usepackage{cleveref}
\usepackage{dsfont} 
\usepackage[ruled]{algorithm2e} 

\usepackage{graphicx}

\usepackage{amsmath}
\usepackage{amsfonts}
\usepackage{amsthm}  
\usepackage{dsfont}  
\usepackage{amsthm}

\newtheorem{theorem}{Theorem}

\makeatletter
\let\saved@includegraphics\includegraphics
\AtBeginDocument{\let\includegraphics\saved@includegraphics}
\renewenvironment*{figure}{\@float{figure}}{\end@float}
\makeatother

\title{Efficient message passing for cascade size distributions on finite trees}

\author{Rebekka Burkholz$^{1}$}

\begin{document}

\maketitle

\begin{affiliations}
\item Department of Computer Science, ETH Zurich, Zurich 8006, Switzerland
\end{affiliations}

\begin{abstract}
How big is the risk that a few initial failures of networked nodes amplify to large cascades that endanger the functioning of the system?  
Common answers refer to the average final cascade size.
Two analytic approaches allow its computation: a) (heterogeneous) mean field approximation and b) belief propagation.
The former applies to (infinitely) large locally tree-like networks, while the latter is exact on finite trees. 
Yet, cascade sizes can have broad and multi-modal distributions that are not well represented by their average.  
Full distribution information is essential to identify likely events and to estimate the tail risk, i.e. the probability of extreme events.
Here, we lay the basis for a general theory to calculate the cascade size distribution in finite networks.  
We present an efficient message passing algorithm that is exact on finite trees and a large class of cascade processes.
An approximation version performs well on locally tree-like networks.
\end{abstract}

Mean field theories are core to the analysis of stochastic processes on networks, as they make them analytically tractable and allow the estimate of average quantities of interest. 
Fundamental to this approach is the configuration model and its variants\cite{Newman.Strogatz.ea2001Randomgraphswith}. 
These create random network ensembles whose locally tree-like network structure is exploited to approximate the average neuronal activity in a brain\cite{brain2,brain}, estimate the size of an epidemic outbreak\cite{NewmanSIR}, measure systemic risk\cite{Battiston2012a}, or analyze the formation of opinions\cite{Watts2002}.  
Their analysis has deepened our understanding of cascade phenomena and provided insights into the average role of connectivity in the spreading of failures or activations\cite{Burkholz2015,BurkholzMultiplex,LRD,BurkholzCorrelations}.  
In consequence, many of our insights rely on Local Tree Approximations (LTA) and, thus, the assumption that large systems can be approximated well by their infinitely large counterpart and that neighbors of the same node are independent. 
Finite systems that are small enough so that finite size effects have to be considered are subject of study in many important applications. 
For a given and fixed network, belief propagation (BP), also termed cavity method in Physics, serves the computation of average node states and thus the average final cascade size.  
Furthermore, it can provide means to estimate the probability of extreme events in large systems\cite{BianconiExtreme}. 
As LTA, BP relies on independent neighbors and is thus exact on trees, while an iterative application (i.e. loopy belief propagation) approximates well average cascade results on locally tree-like networks\cite{BP}.
Yet, finite networks, even when they are large, can behave quite different from the expected, in particular close to phase transitions. 
The distribution of the final cascade size can be broad and even of multi-modal shape as shown for specific topologies, i.e. complete networks and stars\cite{BurkholzFinite}. 
Another example is the well known Curie-Weiss model\cite{friedli_velenik_2017}, whose magnetization density distribution is bi-modal for low temperature. 
Also real world applications elucidate the need for distribution information in addition to averages\cite{ThurnerFinance}. 
Both Local Tree Approximations\cite{BPthresholds} and Belief Propagation\cite{BP} can be formulated as a message passing algorithm, which can be distributed efficiently over several computing units. 
We present a third one. 
Yet, it provides the full cascade size distribution. 
In contrast to BP, we only have to go through a tree once instead of twice. 
As in each node cascade size distributions of subtrees rooted in its children are combined, we term this approach \emph{Subtree Distribution Propagation (SDP)}. 
It is exact on trees and efficient. 
For limited resolution of the cascade size, it only requires a number of operations that is linear in the number of network nodes: $O(N)$. 
To further approximate the cascade size distribution on general networks, we introduce a second algorithm: termed \emph{Tree Distribution Approximation (TDA)}. 
It relies on loopy belief propagation (or another algorithm to compute marginal activation probabilities of nodes) and SDP. 
By comparison with extensive Monte Carlo simulations, we show that TDA approximates the cascade size distribution on locally tree-like networks well.
As we discuss further, our derivations can form the basis of algorithms for general network topologies.

\section*{Cascade model framework}
We assume that a fixed undirected network (or graph) $G = (V,E)$ with node (or vertex) set $V$ and link (or edge) set $E$ is given. 
Each $i \in V$ of the $N = |V|$ nodes is equipped with a binary state $s_i \in \{0,1\}$, where $s_i = 1$ indicates that $i$ is active (or failed) and $s_i = 0$ that $i$ is inactive (or functional). 
In the course of a cascade, node states can become activated by local interactions with network neighbors, i.e. the nodes a node is connected with by links.
Note that activation can travel in both directions of a link. 
We assume that the process evolves over discrete time steps $t = 0, \cdots, T$ and that the activation of a node $i$ at time $t$ depends on the number $a_i(t-1)$ of active neighbors at the previous time step. 
The respective cascade model is defined by the response functions $R_i$ for each node $i \in V$.
A node $i$ activates with probability $R_i(a)$ when exactly $a$ of its neighbors are active (while $a-1$ would not have been enough). 
%
Thus, $i$ activates with probability $R_i(0)$ and never activates with probability $R_i(d_i + 1)$, where $d_i$ denotes $i$'s degree, i.e. the number of its neighbors. 
We further define $R^c_i(a)$ as probability that a node becomes active whenever $a$ neighbors are active.  
Usually, this is the cumulative sum $R^c_i(a) = \sum^a_{l=0} R_i(l)$ and we have $\sum^{d_i+1}_{a=0} R_i(a) = 1$. 
This reflects the reasoning that each active neighbor increases the chance to activate the node. 
For instance in opinion formation models, also opposite effects could be thought of, i.e. a high number of active neighbors reduces the probability of adopting the same opinion.  
For simplicity, we assume that $R_i$ is not time dependent itself and exclude the possibility of recovery, i.e. that a node switches from an active/failed ($s_i = 1$) back to an inactive/functional state ($s_i = 0$).  
In principle, the recovery of a node could be considered by the introduction of a third node state $s_i =$ 'recovered', but would introduce additional computational complexity that we avoid here.  
In this setting, we are interested in the final cascade size that is measured by he final fraction of active nodes $\rho = \frac{1}{N} \sum^{N}_{i=1} s_i(T)$.
It answers, for instance, the question how many nodes receive a certain information or how many pass on a disease. 
Regardless whether we want to minimize or maximize $\rho$, considering the probability of adverse events can improve the decision making. 
This framework covers many cascade models, ranging from neural dynamics to Voter models\cite{GleesonPairApproximation,wattsCollect}. 
Two common examples shall be discussed in more detail: 
(a) a threshold model (TM) of information propagation\cite{Granovetter1978,Watts2002}. and (b) a simple model of epidemic spreading, also termed independent cascade model (ICM)\cite{Kermack700,NewmanSIR,Kempe2003}. 
Details are provided in the method section. 

\begin{figure}[htpb]
  \centering
  \subcaptionbox{Subtree Distribution Propagation.
  \label{fig:SDP}}{
     \includegraphics[width=0.4\textwidth]{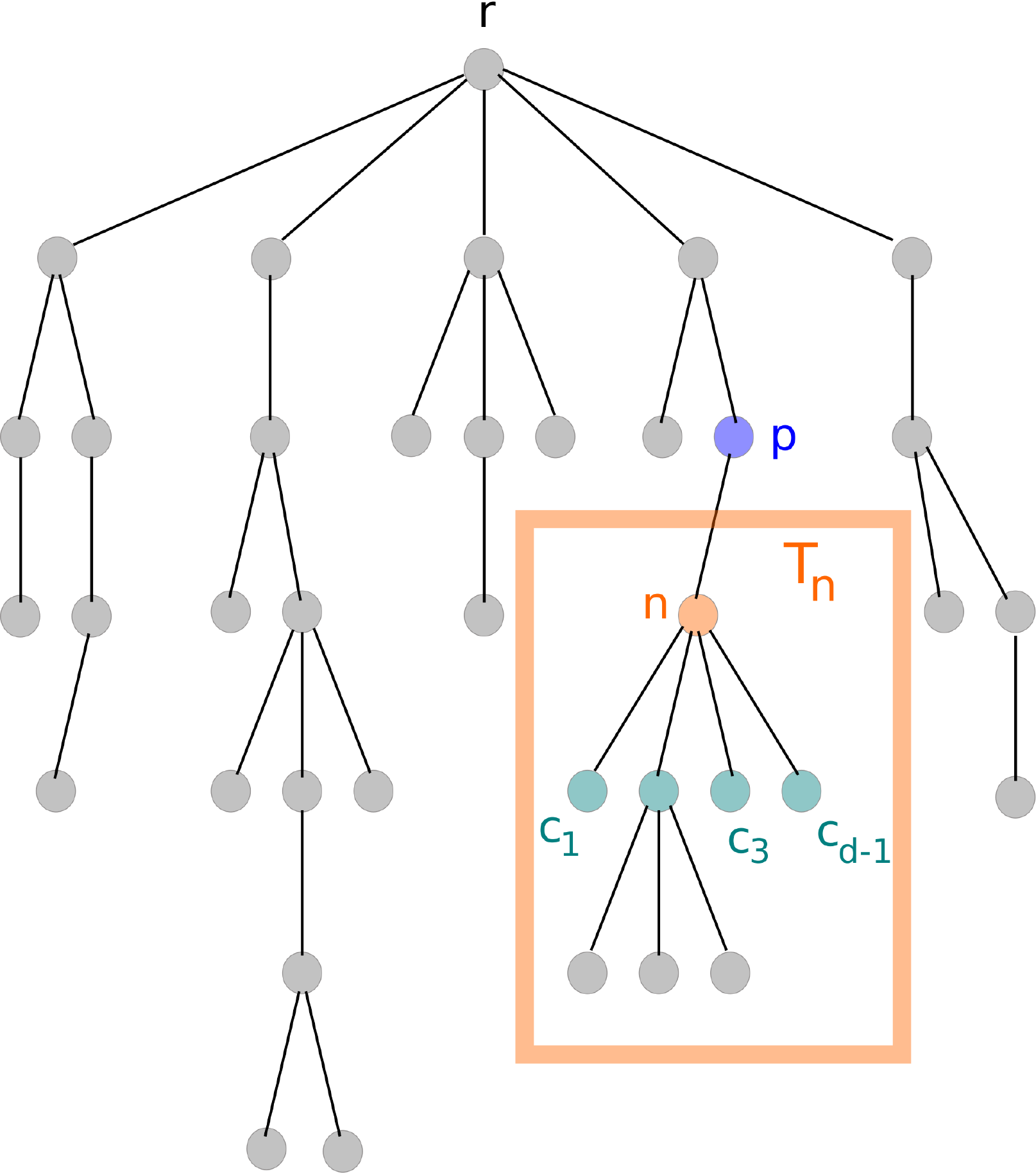}
  }
  \subcaptionbox{Tree Distribution Approximation.
  \label{fig:TDA}}{
     \includegraphics[width=0.4\textwidth]{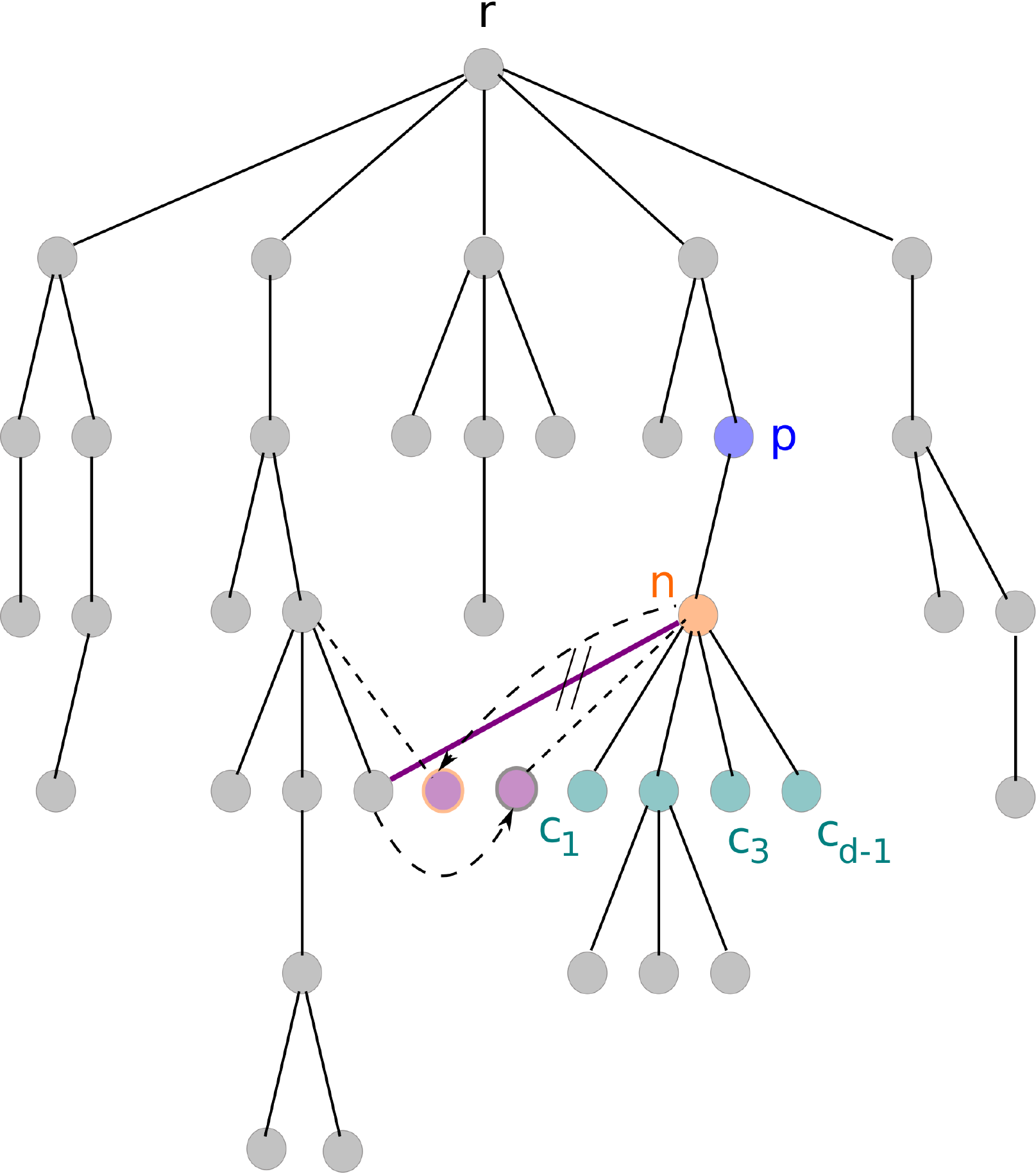}
  }
 \caption{Illustration of relevant variables in the message passing algorithm. $n$ denotes a focal node, $p$ its parent, and $T_n$ the subtree rooted in $n$. The calculation starts in the leaves (the bottom nodes with degree $1$) and successively computes the cascade size distribution of each subtree $T_n$ given the state of the parent $p$ by combining the distributions corresponding to trees rooted in the children $c_i$. The resulting distribution is exact for a tree (a). If the network contains loops (b), (purple) links are deleted until a tree is obtained. Each deleted link is replaced by two new links that reconnect a independent (purple) copy of a cut-off neighbor. Such a copy is not counted as additional node in the final cascade size, but influences the activation probability of its neighbor $n$.}
  \label{fig:algoExplain}
\end{figure}

\section*{Subtree distribution propagation (SDP)}

Fig.~\ref{fig:SDP} visualizes the general procedure of our exact message passing algorithm to calculate the final cascade size distribution for a tree with root $r$.
We call it subtree distribution propagation (SDP), as it is based on the idea to calculate the cascade size distribution for each subtree $T_n$ rooted in a node $n$ given the state of its parent $p$. 
We start in the leaves (i.e. the nodes with degree $1$) at the highest level (i.e. at the bottom of the picture) and proceed iteratively upwards to the root $r$ by combining the subtree distributions corresponding to the children. 
In slight abuse of notation, let $T_n$ denote the number of active nodes in the subtree rooted in $n$, which we also call subtree cascade size (as $T_n/N$).
This is a random variable that can be expressed as sum over the node state $s_n$ and children subtrees: $T_n = s_n + \sum^{d_n-1}_{i=1} T_{c_i}$.
$T_n$ and all involved node states depend on $s_p$ (and each other) in complicated ways. 
We control for this dependency by introducing an order-conditioning operator $\parallel$ that has a similar function as conditioning on random variables. 
Yet, exact conditioning $T_n \mid s_p = 1$ would consider events where $n$ causes the activation of $p$ and vice versa. 
However, we have to take care of the right order of activations. 
$T_n \parallel s_p$ denotes the cascade size of a tree $T_n$ where the rest of the original network has been removed and $n$ has an additional neighbor $p$, whose state is set to $s_p$ with probability $1$. 
This way, we forget about the influence of $n$ on $p$ (at this point). 
Computing the distribution of $T_n \parallel s_p$ is challenging for two reasons: a) the random variables are dependent and b) the right order of activations needs to be respected.
The solution for a) is to order-condition $T_n$ on events involving $s_n$ (and $s_p$) that make the subtree distributions independent so that $T_n$ is given by their convolution.
Convolutions can be computed efficiently with the help of Fast Fourier Transformations (FFTs).
To solve b), we define artificial variables $I_n$, $A_n$ that capture the right order of activations and the dependence structure of $s_n$ on $s_p$ and $T_{c_i}$.
$I_n$ refers to an inactive and $A_n$ to an active parent $p$.  
Their distributions $p_{I_n}$, $p_{A_n}$ are advanced iteratively so that we can assume their knowledge for the children $I_{c_i}$, $A_{c_i}$. 
Combined, they add the subtree cascade sizes and, separately, the number of active children $a_n$ that can trigger the activation of $n$.  
Thus, in our subtree distribution propagation algorithm, each node (except the root) sends exactly one message to its parent: the distribution of $I_n$ and $A_n$ (or better: their Fourier transform).  
This message is a combination and update of the messages the node received by its children, which is detailed in the method section.  
The root finally combines all received messages to compute the final cascade size distribution $p_{\rho}(x) = \mathbb{P}\left(T_r = x N\right)$.
All of this is very fast for limited resolution of the cascade size. i.e. when we restrict $\rho$ on an equidistant grid of ${[0,1]}$. 
Then, the algorithmic complexity of SDP is linear in the number of nodes: $O(N)$.
It can further be brought down to $O(h)$, where $h$ denotes the height of the tree, if the computations are distributed to computing units corresponding to nodes of the tree.  
A detailed analysis is provided in the Supplementary Information.

\section*{Tree distribution approximation (TDA)}

SDP is exact on trees. 
However, activations are stronger coupled in the presence of loops and the probability of large and small cascades tends to increase\cite{BurkholzFinite} so that the variance of the cascade size distribution grows.
To take this into account, we propose an approximation version of SDP. 
The idea is to first calculate individual activation probabilities on the original network and second to use them for adapting the response functions $R_i$. 
These are given as input to SDP which is applied to a minimum spanning tree of the original network.  
Since this approach is only approximate and is based on the cascade size distribution on a tree, we call it tree distribution approximation (TDA). 
In detail, we employ loopy BP to calculate the activation probabilities $p_{in} = \mathbb{P}(s_i = 1 \parallel s_n = 0)$ of a neighbor $i$ given that $n$ is not active (before) to update the response function $R_n$, as outlined in the method section. 
Loopy BP itself is not exact, yet, usually approximates $p_{in}$ well on locally tree-like networks. 
It could be substituted by any alternative algorithm. 
For instance, the Junction Tree Algorithm \cite{Jordan} would be exact but computationally costly and does not scale to large networks. 
Next, we compute a minimum spanning tree of the original network (i.e. delete links of loops until we obtain a tree). 
Further, we assume that lost neighbors $i$ of a node $n$ activate initially and independently (before $n$) with probability $p_{in}$ so that they can still contribute to the activation of $n$. 
Fig.~\ref{fig:TDA} illustrates this approach. 
We create an independent copy of a lost neighbor $i$ (which is colored purple) and connect it with $n$.
The copy's activation is not counted in the final cascade size $\rho$. 
It only influences the response $R_n$. 
Therefore, this algorithm neglects certain dependencies of node activations in the presence of loops. 
If these loops are large enough, their contribution is usually negligible.
Therefore, we expect to approximate cascade size distributions well on locally tree-like networks. 
Next, we test this claim in numerical experiments. 

\section*{Numerical Experiments}
\begin{figure}[htpb]
  \centering
  \subcaptionbox{Tree. 
      \label{fig:tree}}[.45\linewidth]{%
     \includegraphics[width=0.3\textwidth, scale=0.1]{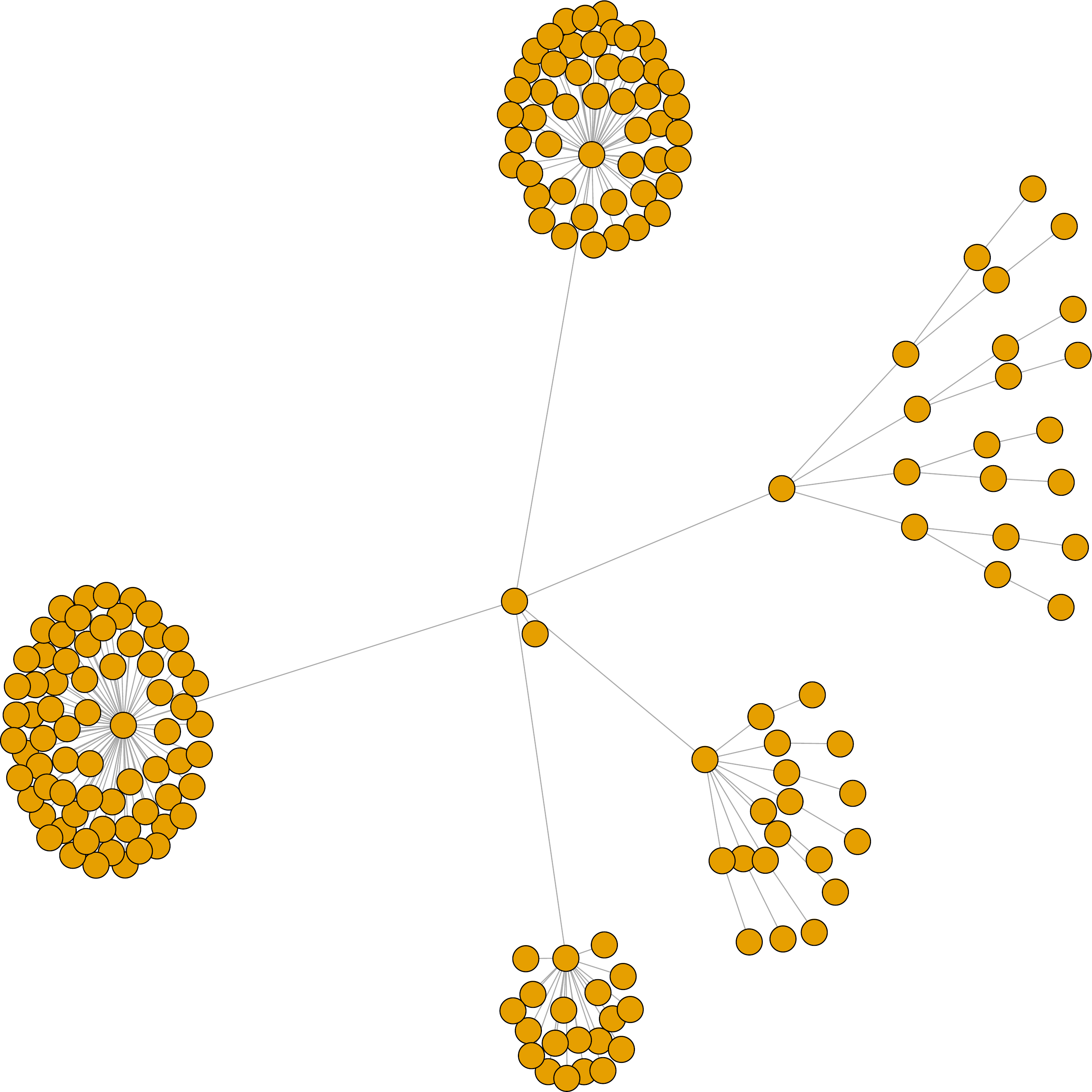}
  }
    \subcaptionbox{Tree: SDP.
      \label{fig:treeDistr}}[.45\linewidth]{%
     \includegraphics[width=0.43\textwidth]{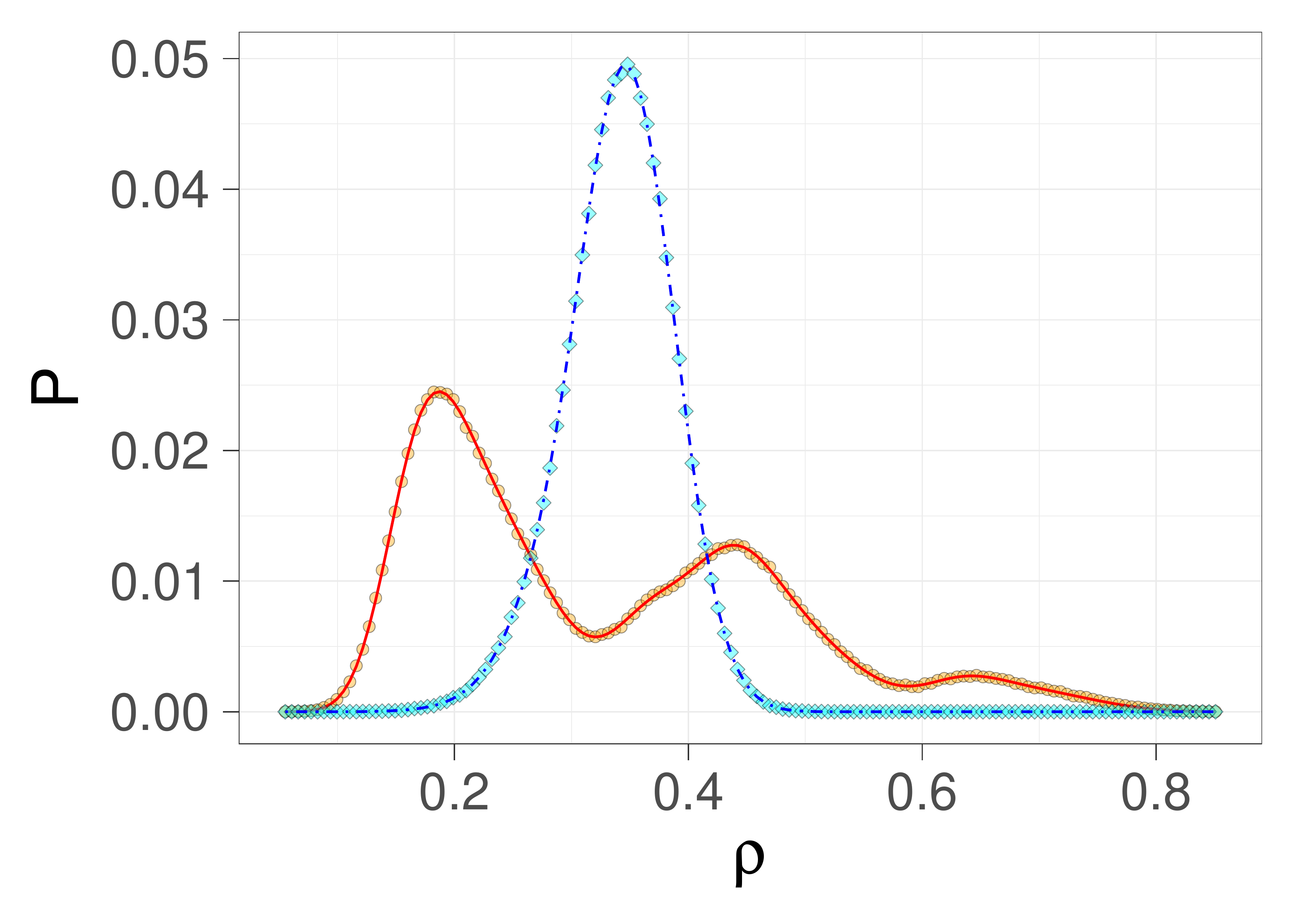}
  }
  \subcaptionbox{Configuration model network.
  \label{fig:config}}[.45\linewidth]{%
     \includegraphics[width=0.35\textwidth]{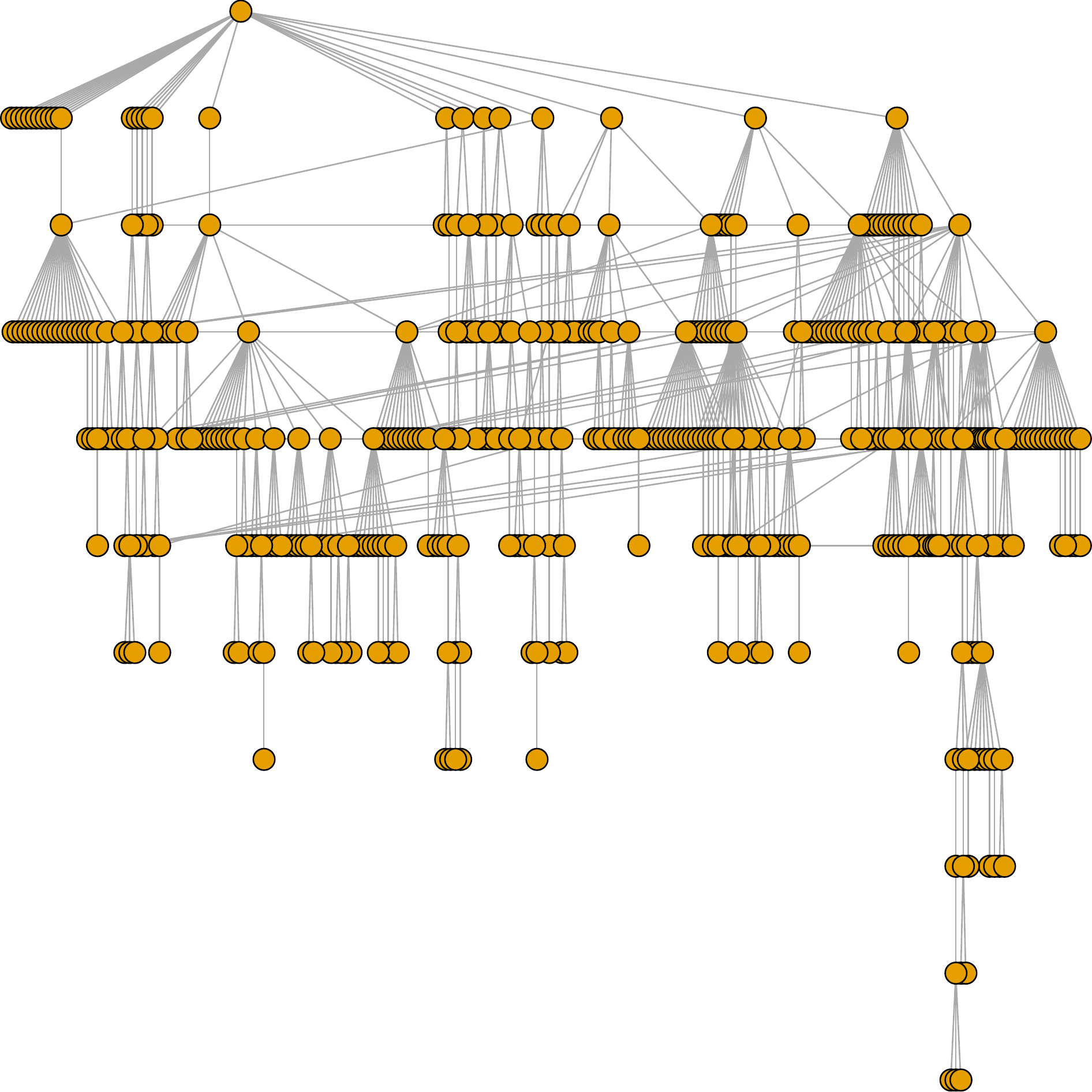}
  }
    \subcaptionbox{Configuration model network: TDA.
  \label{fig:configDistr}}[.45\linewidth]{%
     \includegraphics[width=0.43\textwidth]{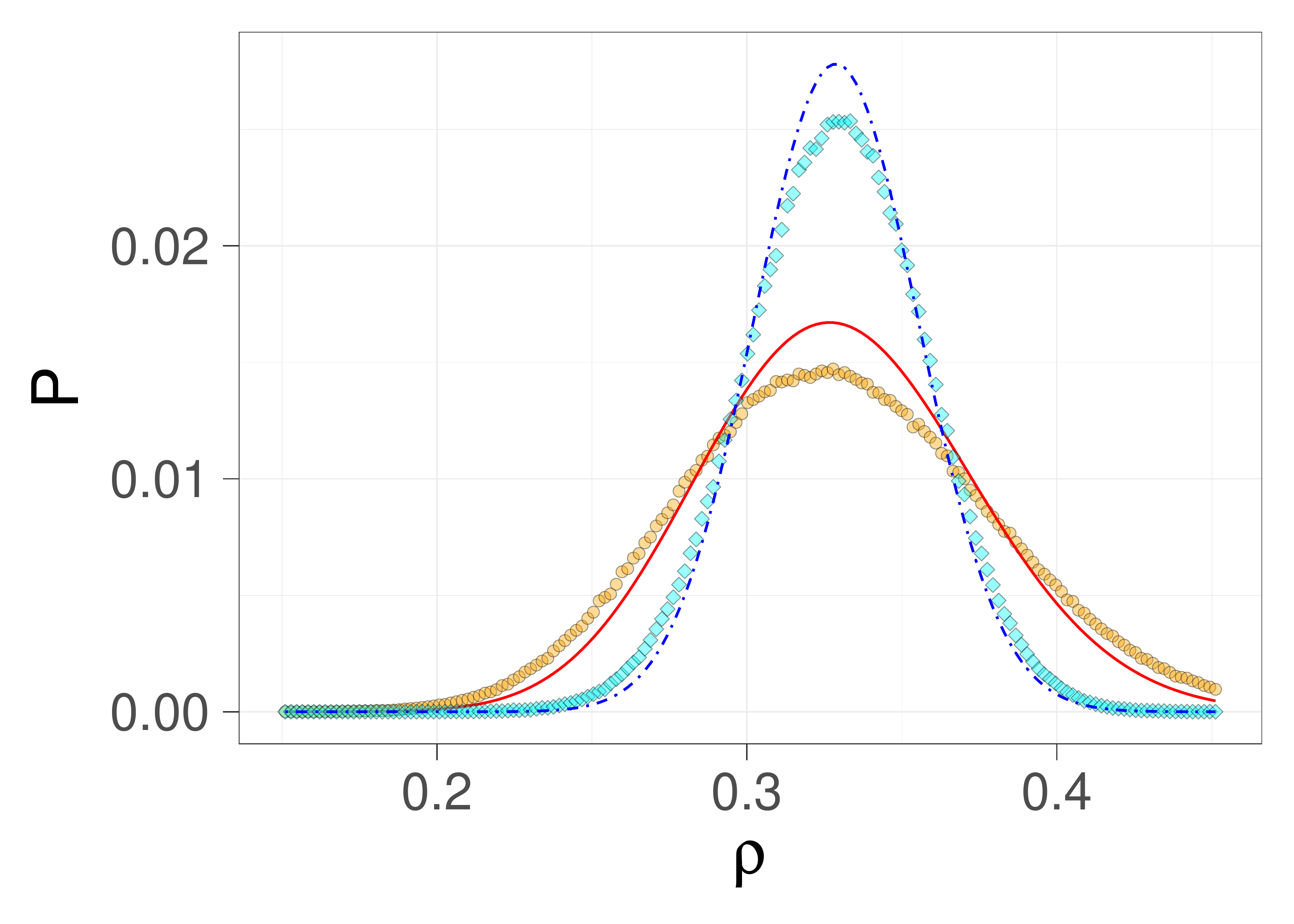}
  }
  \subcaptionbox{Corporate ownership network. 
  \label{fig:EVA}}[.45\linewidth]{
     \includegraphics[width=0.3\textwidth]{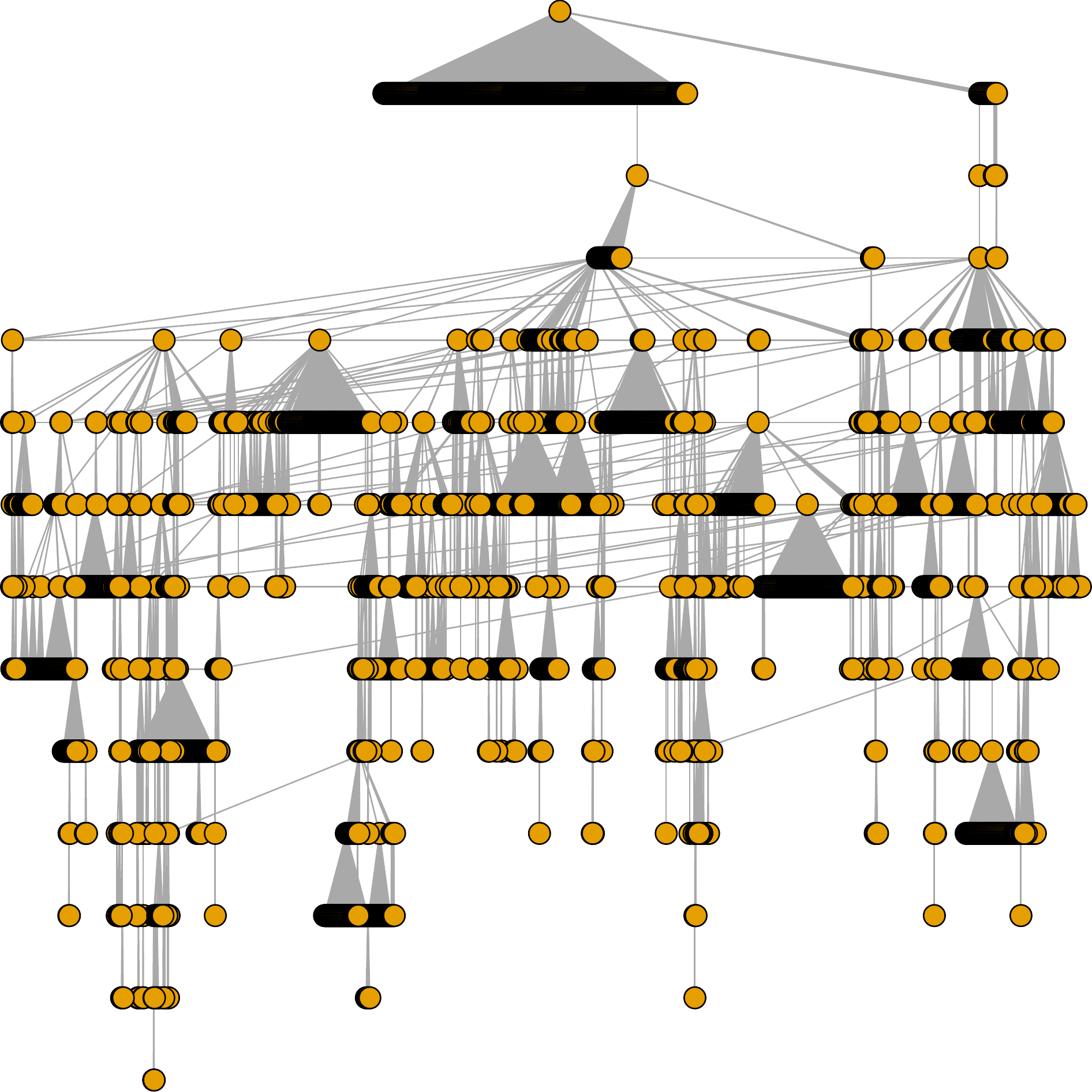}
  }
  \subcaptionbox{Corporate ownership network: TDA.
  \label{fig:EVADistr}}[.45\linewidth]{
     \includegraphics[width=0.43\textwidth]{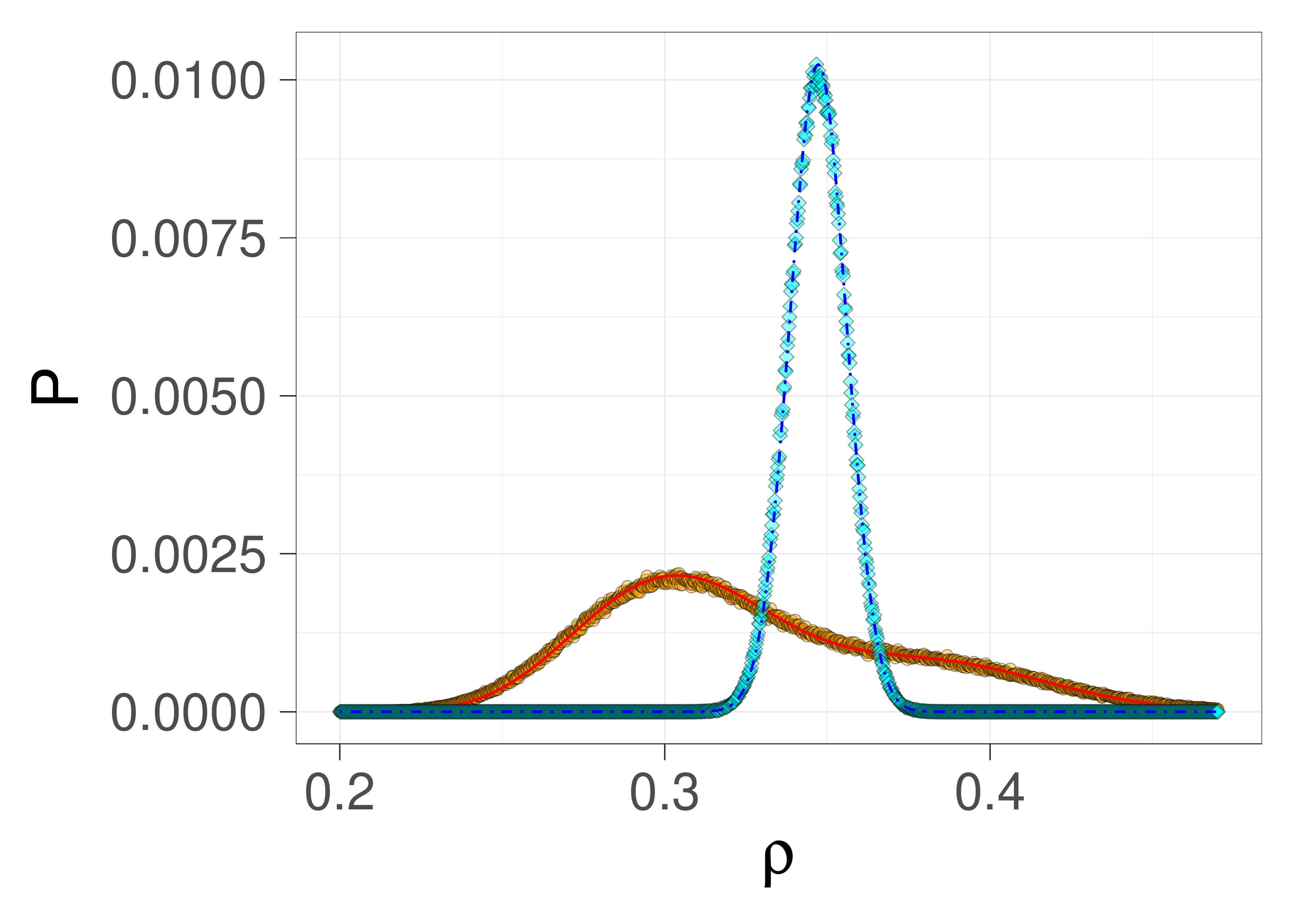}
  }
  \caption{Cascade size distribution on exemplary networks. The left column shows the network, the right column the corresponding cascade size distributions. 
  }
  \label{fig:exp}
\end{figure}
\addtocounter{figure}{-1}
\begin{figure}[t]
 \caption{Symbols represent Monte Carlo simulations (with $10^6$ realizations): orange circles for the threshold model and cyan squares for the independent cascade model. Lines correspond to the respective message passing algorithm: a solid red line represents the threshold model and a blue dotted line the independent cascade model.
 }
\end{figure}
We focus on three exemplary networks that are representative of different use cases and visualized in Fig.~\ref{fig:exp}: a tree, a locally tree-like network constructed by a configuration model with power law degree distribution, and a real world network defined by data on corporate ownership relationships\cite{EVAData}, which is is locally tree-like.
For each network, we compare the cascade size distributions obtained by our message passing algorithm, i.e. SDP for the tree and TDA for the two other networks, with Monte Carlo simulations. 
We focus on the two introduced cascade models with the same parameter setting for all networks as specified in the method section. 
This provides a proof of concept and allows to assess the approximation quality of TDA in Fig.~\ref{fig:exp}. 
First, we observe that SDP and TDA match perfectly the cascade size distributions obtained by extensive Monte Carlo simulations for the tree and the locally tree-like corporate ownership network.
For the power law configuration model, where the task is much harder, TDA identifies the modes correctly, yet, tends to slightly underestimate the variance of the cascade size distribution.
A considerable number of loops introduces additional correlations of note states that we cannot capture by our tree approximation. 
Still, we provide a slightly improved estimate of the average cascade size over BP and a lower bound for the variance.
Second, we note the broad cascade size distributions. 
This is unexpected by heterogeneous mean field or BP analysis, as our parameter choices for the cascade models are in no case critical: Neither does the average cascade size undergo a phase transition close to the chosen parameters in an infinitely large network with the same degree distribution as the original network, nor does the average cascade size change abruptly in the finite network for small changes in the parameters. 
For the threshold model, we also observe several modes of the distribution on the tree and corporate ownership network.
Clearly, the average cascade size does not represent the cascade risk well in these cases. 
Our approaches, SDP and TDA, add cascade size distribution information. 
These are useful in particular when we face star structures or, similarly, pronounced hubs (i.e. nodes with large degree), as these contribute to multiple distribution modes. 
The modes roughly correspond to events where no hub activates, one hub activates (so that many of its neighbors follow), two activate, etc., while longer paths have a smoothing effect on the distribution. 
The independent cascade model shows single modes only, since we analyze parameters here where it is very likely that the center becomes active but does not substantially increase the activation probability of its neighbors. 
Apriori, the precise shape of the cascade size distribution for complicated network structures is not clear and calls for a detailed analysis with the provided tools. 

\section*{Discussion}
We have introduced two algorithms that compute the final cascade size distribution for a large class of cascade models: a) the subtree distribution propagation (SDP) is exact on trees, while b) the tree distribution approximation (TDA) provides an approximation variant that performs well for locally tree-like network structures. 
Their derivation is based on two basic ingredients: artificial random variables that consider the right order of activations and an order-conditioning operation where the network above a node's parents are cut off. 
The latter creates an independence of subtree cascade size distributions, which enables their efficient combination. 
For limited resolution of the cascade size distribution, the SDP part of the algorithms is linear in the number of nodes $O(N)$ and can be distributed along the tree structure of the input.
Each node needs to be visited only once. 
In consequence, the introduced algorithms are quite efficient and scalable. 
As we argue, cascade size distribution information is critical for good decision making, when the distributions are broad and, in particular, when they have multiple modes, which signify probable events.
Therefore, there is a need to generalize our approach beyond locally tree-like network structures, i.e. to networks with higher loop density.  
This generality will trade off with efficiency and scalability, similarly as the junction tree algorithm relates to belief propagation. 
The approach presented here lends itself as well for a transfer to junction trees. 
On a meta level, we have presented a way to combine cascade size distributions of subnetworks and do not rely on the assumption that these subnetworks are trees themselves.
Their distribution can either be computed analytically or approximated by Monte Carlo simulations.  
In every case, we can efficiently combine the related distributions if the subnetworks are connected in a tree-like fashion (as in junction trees).
Furthermore, the principle of our approach can be transferred to more general graphical models to obtain macro level information as, for instance, the distribution of the sum of involved random variables. 

\begin{methods}
\subsection{Cascade models}
We analyze two models in more detail, termed threshold model (TM) and independent cascade model (ICM).  
Both models have been used to describe similar phenomena, as information propagation, opinion formation, social influence, but also financial contagion or the spread of epidemics.  
While the cascade mechanisms are similar for both, an important distinction is that in the threshold model the probability to activate a neighbor depends on the other activations of neighbors\cite{wattsCollect}.

\paragraph{The threshold model} originates in a model of collective action\cite{Granovetter1978}, which has been transferred to networks by\cite{Watts2002}.
Each node $i$ is equipped with a threshold $\theta_i$ that has been drawn initially independently at random from a distribution with cumulative distribution function $F_i$. 
Nodes with a negative threshold become active initially. 
Otherwise, a node activates whenever the fraction of active neighbors exceeds its threshold, i.e. $\theta_i \leq a/d_i$. 
In consequence, previous activations of neighbors influence the probability whether a further activation of a neighbor causes the activation of the focal node $i$. 
This implies a response function of the form:
\begin{align*}
\begin{split}
& R_i(0) = F_i(0), R_i(a) = F_i\left(\frac{a}{d_i}\right)  -  F_i\left(\frac{a-1}{d_i}\right) ,   R_i(d_i + 1) =  1 -  F_i\left(1\right), R^c_i(a) =  F_i\left(\frac{a}{d_i}\right). 
\end{split}
\end{align*}

\paragraph{The Independent cascade model} can be interpreted as simple epidemic spreading model that resembles the widely studied SIR (Susceptible-Infected-Recovered) model\cite{Kermack700}. 
It is also equivalent to bond percolation in terms of the final outcome\cite{NewmanSIR}. 
The activation of a node corresponds to its infection. 
Even though we do not explicitly allow for node recovery, for large networks, it can be implicitly incorporated in the choice of the infection probability $p$, i.e. the probability that a newly infected (active) node spreads a disease to a network neighbor. 
All neighbors of a newly infected node are infected independently.
Also initially, nodes are activated independently with probability $p$. 
Thus, a node with degree $d_i$ has the response function:
\begin{align*}
\begin{split}
& R_i(a) = p*(1-p)^a, \ \ \  R_i(d_i + 1) =  (1-p)^{d_i+1},  \ \ \  R^c_i(a) = 1 - (1-p)^{a+1}
\end{split}
\end{align*}
for $ 0 < a \leq d_i$.
A node becomes activated exactly with $a$ active neighbors (if it is not active initially, which is the case with probability $1-p$) and one out of the $a$ neighbors causes the activation with probability $p$, while the remaining $a-1$ did not cause the activation.

Average cascade properties have been extensively studied for both models with the help of heterogeneous mean field approximations (for TM\cite{Gleeson2007,BurkholzMultiplex,Burkholz2015,BurkholzCorrelations,LRD} , for ICM\cite{NewmanSIR}) and belief propagation (for TM\cite{BPthresholds}, for ICM and more complicated variants\cite{PhysRevX.4.021024,DBLP:conf/nips/Lokhov16}).


\paragraph{Numerical experiments}
We run experiments for three networks that are visualized in Fig.~\ref{fig:exp}.
A tree and the configuration model network are created artificially, while the last one is a real world example based on data. 
The tree consists of $N=181$ nodes with two main hubs of degrees $69$ and $50$, while the configuration model network is a bit larger with $N=543$ nodes and average degree $d_{\rm{avg}} = 2.25$, but smaller maximal degree $d_{\rm{max}} = 25$.
The latter has degree distribution $p(d) \propto d^{-2.5}$, which is structurally close to many real world networks \cite{Barabasi1999,Clauset2009}.
While the configuration model constructs locally tree-like networks, the size $N=543$ is chosen on purpose relatively small so that the network has still a number of short loops as visible in Fig.~\ref{fig:config}. 
This makes our approximation task harder and serves as stress test for our approach.
The largest considered network is the largest weakly connected component of a publicly available network, which is defined by corporate ownership relationships\cite{EVAData}. 
It consists of $|V| = 4475$ nodes with mean degree $z = 2.08$ and maximal degree $d_{max} = 552$ and is clearly locally tree-like. 

We compare the final cascade size distribution given by our algorithms with the results of Monte Carlo simulations always for the two introduced cascade models with the same parameter setting.
In the threshold model, we assume independently normally distributed thresholds with a given mean $\mu =0.5$ and standard deviation $\sigma =0.5$ so that $F_i(\theta) = \Phi\left((x-\mu)/\sigma\right)$ for all $i \in V$, where $\Phi$ denotes the standard normal cumulative distribution function.
The parameter $p$ in the independent cascade model is always set to $p = 0.2$. 
This parameter choice is non-critical and thus, no phase transitions occur in close neighborhood of the parameters. 
For Monte Carlo simulations, we always report the empirical distribution of $10^6$ independent realizations.  
We calculate the final cascade size distribution for the tree by SDP and for the other two locally tree-like networks by TDA at full resolution, i.e. $\rho \in \{0,1/N, 2/N, ... ,1\}$.

\subsection{Subtree distribution propagation}

The goal is to compute the final cascade size distribution $p_{\rho}(t/N) = \mathbb{P}\left(T_r = t \right)$ for a given tree $G=T_r$ with root $r$ and cascade model with response functions $R_i$ by a message passing algorithm. 
As explained in the main text, nodes $n$ send messages $p_{I_n}$, $p_{A_n}$ to their parent $p$, where $I_n$ refers to an inactive ($s_p =0$) and $A_n$ to an active parent ($s_p = 1$). 
\begin{figure}[htpb]
  \centering
  \subcaptionbox{Subtree snippet.  
  \label{fig:treeZoom}}{
     \includegraphics[width=0.3\textwidth]{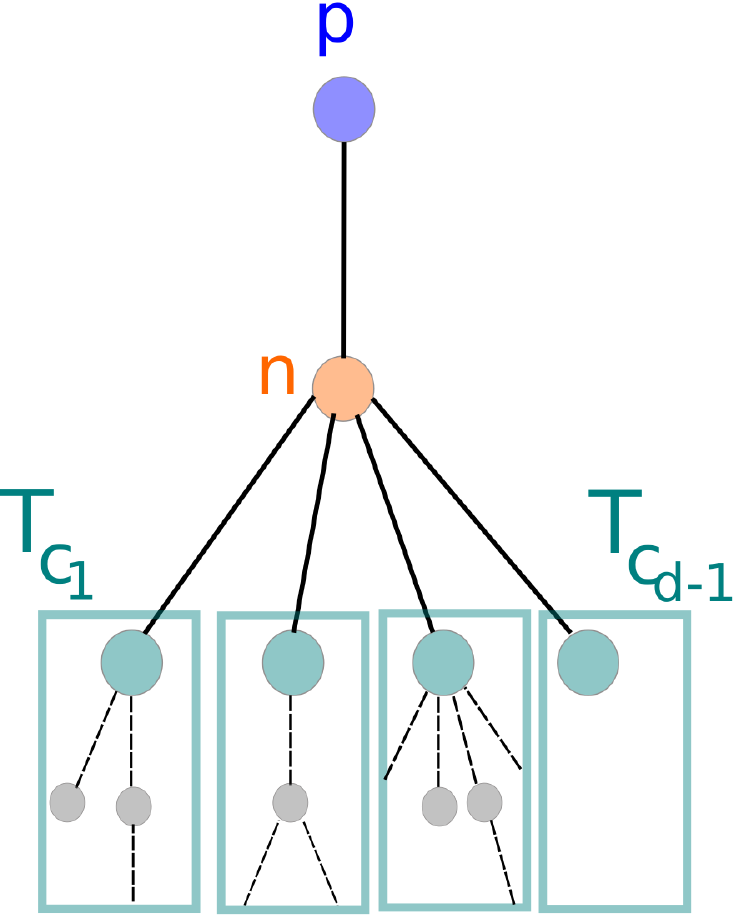}
     \hspace*{\fill}
  }
      \subcaptionbox{Before activation of $n$. 
      \label{fig:treeZoom1}}{%
     \includegraphics[width=0.3\textwidth]{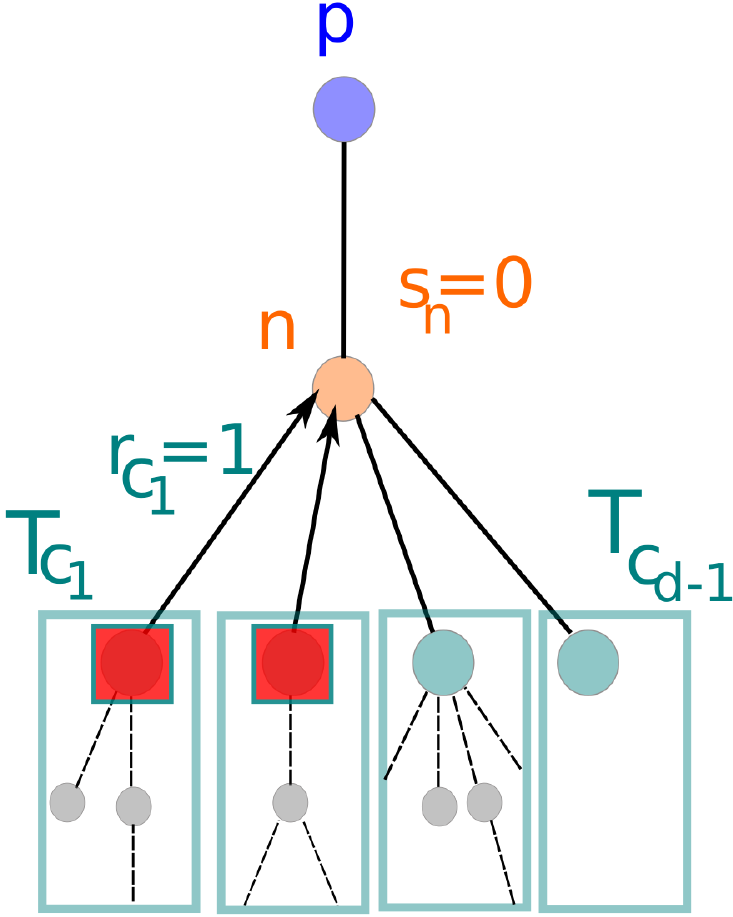}
  }
  \subcaptionbox{After activation of $n$.
  \label{fig:treeZoom2}}{
  \hspace*{\fill}
     \includegraphics[width=0.3\textwidth]{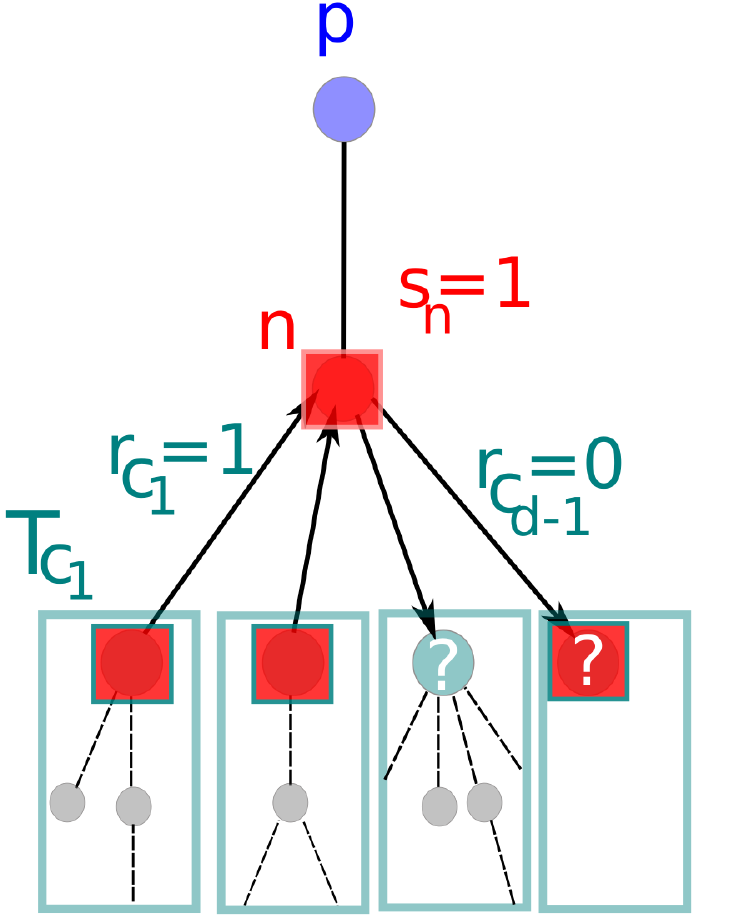}
  }
  \caption{Illustration of artificial state $r_{c_i}$. $n$ denotes the focal node with degree $d$, $p$ its parent, and $c_1, ..., c_{d-1}$ its children. $T_{c_i}$ is a subtree rooted in a child $c_i$. An active node is represented by a red square. $r_{c_i}$ denotes an artificial state of child $c_i$ that indicates with $r_{c_i} = 1$ whether (b) it became active before its parent $n$ and can thus trigger its activation or with $r_{c_i} = 0$ whether (c) it activates after its parent $n$ or not at all.}
  \label{fig:treeExplain}
\end{figure}
To explain how the definition of $I_n$ and $A_n$ is useful, we shift the focus from $n$ to its children and show how the messages corresponding to them enable us to compute the distribution of $T_n = s_n + \sum^{d_n-1}_{i=1} T_{c_i}$. 
Let's first discuss the easier case when $n$ stays inactive. 
The subtree distributions of $T_{c_i} \parallel s_n = 0$ are independent.
Thus, we can convolute the distributions of $I_{c_i} = \left(T_{c_i}, s_{c_i}\right) \parallel s_n = 0$ to obtain the distribution of $(T_n, a_n) \parallel s_n = 0$ with $a_n =  \sum^{d_n-1}_{i=1} s_{c_i}$. 
In this case, we know the probability that $s_n$ does not become active (given its parent $p$): $\mathbb{P}\left(s_n = 0 \parallel a_n, s_p\right) = 1- R^c_n(a_n+s_p)$. 
The case $s_n = 1$ is more involved, since we have to consider only the children that trigger the activation of $n$, i.e. that become active before $n$.
We therefore introduce an artificial binary node state $r_n$, which is illustrated by Fig~\ref{fig:treeExplain}. 
$r_n = 1$ indicates that node $n$ activates before its parent $p$ and contributes to its activation, while $r_n = 0$ subsumes all other cases leading to $s_p = 1$, i.e. $n$ does not activate before its parent, has an active parent, and might become active or not after the activation of its parent. 
We join $r_n$ with an adapted subtree cascade size $\tilde{T}_n$ to $A_n = (\tilde{T}_n, r_n)$ so that $\sum^{d_n-1}_{i=1} A_{c_i} = \left(T_n-1,  a_n \right) \parallel s_n = 1$ with now $a_n = \sum^{d_n-1}_{i=1} r_{c_i}$. 
$\tilde{T}_n$ depends on $r_n$ and $s_n$ and is defined as $\tilde{T}_n \ = \  r_n T_n  \mathds{1}_{ \{ s_n = 1 \} }   \; \parallel \; s_p = 0 \ \  + \  (1-r_n) T_n    \mathds{1}_{\{s_p = 1 \rightarrow s_n = 1  \ \lor \ s_n = 0\}} \; \parallel \; s_p = 1$. 
Thus, if $r_n = 1$, $n$ is active ($s_n = 1$) and $T_n$ is not influenced by its parent, i.e. $s_p = 0$ is given. 
If $r_n = 0$, the parent is assumed to be active $s_p = 1$ and the node itself can either be inactive $s_n = 0$ or, if it activates ($s_n = 1$), $p$ contributes to its activation so that $n$ did not become active before $p$. 
Technically, $A_n$ is not a random variable, since it is not normalized. 
Yet, its convolution still counts the right cases, which are input to the subtree cascade size distribution for active node $n$ given its parent: $\mathbb{P}(T_n, s_n=1 \parallel s_p)$.
In summary, the SDP starts in the bottom of a tree and computes messages $p_{I_n}$, $p_{A_n}$ in each node, sends them (or their Fourier transform) to the parent $p$ until the final cascade size distribution can be computed in the root.
We make this reasoning explicit with the following theorem. 
\begin{theorem}\label{theorem}
Let $G = (V,E)$ be a tree and $R_i, R^c_i$ for $i \in V$ response functions defining a cascade model. 
The final cascade size distribution $p_{\rho}(t/N) = \mathbb{P}\left(T_r = t \right)$ is given by the result of a message passing algorithm ending in the root $r$, where at each node $n \in V$, the following computations are performed based on $p_{A_{c_i}}, p_{I_{c_i}} $ received from their children:\\ 
Case $d_n = 1$ (leaves):
 \begin{align}\label{eq:leaves}
 \begin{split}
 &  p_{A_n}(0,0) = \mathbb{P}\left(T_n = 0 \parallel  s_p = 1 \right) = \mathbb{P}\left(s_n = 0 \parallel  s_p = 1 \right) = 1 - R_n\left(1\right)\\ 
 &  p_{A_n}(1,0) = \mathbb{P}\left(T_n = 1 \parallel  s_p = 1 \right) = R_n\left(1\right) \\
  & p_{A_n}(0,1) = \mathbb{P}\left(T_n = 0, s_n = 1 \parallel  s_p = 0 \right) = 0 \\
  & p_{A_n}(1,1) = \mathbb{P}\left(T_n = 1, s_n = 1 \parallel  s_p = 0 \right) =  R_n\left(0\right)\\ 
  & p_{I_n}(1,0) = p_{I_n}(0,1) = 0 \\
  & p_{I_n}(0,0)  = \mathbb{P}\left(T_n = 0, s_n = 0 \parallel  s_p = 0 \right) = 1- R(0)\\
   & p_{I_n}(1,1) = \mathbb{P}\left(T_n = 1, s_n = 1 \parallel  s_p = 0 \right) =  R(0).
   \end{split}
 \end{align}
A node with degree $d_n > 1$ receives as input the distributions $p_{A_{c_i}}$,  $p_{I_{c_i}}$ corresponding to its children. 
We define $p_{A_n*}$ and $p_{I_n*}$ as their 2-dimensional convolutions:
\begin{align*}
 & p_{A_n*} (t, f) := p_{A_{c_1}} ** p_{A_{c_2}} ** \cdots ** p_{A_{c_{d_n-1}}}  [t,f] \\
 & p_{I_n*} (t, f) := p_{I_{c_1}} ** p_{I_{c_2}} ** \cdots ** p_{I_{c_{d_n-1}}}  [t,f]. 
\end{align*}
Note that we have  $p_{A_n*} (t, a) = p_{I_n*} (t, a) = 0$ for $t < a$.\\
Case $d_n > 1$, $n \neq r$:
\begin{align}\label{eq:node}
\begin{split}
 &  p_{A_n}(t,0)  =  \mathbb{P}\left(T_n = t , s_n = 0\parallel  s_p = 1 \right) \\
 & + \mathbb{P}\left(T_n = t ; s_n = 1;  s_p=1 \rightarrow s_n=1 \parallel  s_p = 1 \right) \\
 & = \sum^{d_n-1}_{a=0}  p_{I_n*} (t, a) \left(1 - R^c_n\left(a+1\right)\right) + \sum^{d_n-1}_{a=0}  p_{A_n*} (t-1, a) R_n\left(a+1\right)\\ 
  & p_{A_n}(t,1) = \mathbb{P}\left(T_n = t, s_n = 1 \parallel  s_p = 0 \right) =   \sum^{d_n-1}_{a=0}  p_{A_n*} (t-1, a) R^c_n\left(a\right) \\
 & p_{I_n}(t,0)  =  \mathbb{P}\left(T_n = t , s_n = 0\parallel  s_p = 0 \right) = \sum^{d_n-1}_{a=0}  p_{I_n*}(t, a) \left(1 - R^c_n\left(a\right)\right)\\
 & p_{I_n}(t,1)  =  p_{A_n}(t,1), 
  \end{split}
\end{align}
At root $r$: 
\begin{align}\label{eq:root}
\begin{split}
 \mathbb{P}\left(T_r = t\right) = & \mathbb{P}\left(T_r = t, s_r = 0\right)  + \mathbb{P}\left(T_r = t, s_r = 1\right)  \\
 = & \sum^{d_r}_{a=0}  p_{I_r*} (t, a) \left(1-R^c_r(a)\right) + \sum^{d_r}_{a=0}  p_{A_r*} (t-1, a) R^c_r(a) 
 \end{split} 
\end{align}
\end{theorem}
The proof of this theorem is given in the supplementary information along with a pseudocode of SDP.

\paragraph{Algorithmic complexity of SDP.} 
A detailed discussion of the algorithmic complexity of SDP is provided in the Supplementary Information.
In summary, computing the messages $p_{I_n}$ and $p_{A_n}$ in a node $n$ requires $O\left(d_n |T_n| +  |T_n| \log(|T_n|))\right)$ computations, where $|T_n|$ denotes the number of nodes in the subtree rooted in $n$.
Thus, in total $O(\sum^N_{n=1} \left(d_n |T_n| +  |T_n| \log(|T_n|))\right)$ computations are needed to obtain the final cascade size distribution.
Yet, we have two options to reduce the run time: a) limit the accuracy of the cascade size distribution so that $|T_n|$ can be substituted by a constant $C$. For instance, $p_{\rho}$ can be defined only on an equidistant grid of ${[0,1]}$.
In this case, we are left with $O(\sum^N_{n=1} d_n ) = O(N)$ computations. 
b) We can parallelize the matrix times vector multiplications, the Fast Fourier Transformations, and distribute the computations of messages for distinct nodes that are in different subtrees. 
A combination of a) and b) usually leads to an algorithm with smaller run time than $O(N)$, i.e. $O(h)$, where $h$ refers to the height of a tree. 
In the worst case (for instance a long line), this can still require $O(N)$ computations.

Note that the choice of root is relevant for the run time of the algorithm.
Minimizing the maximum path length from the root to any other node in the tree is beneficial in case that enough computing units are available for distribution of the work load. 
In addition, it can be advantageous to place nodes with high degree close to the root so that subtrees are kept small in the beginning. 
Convolutions related to those subtrees operate on small cascade sizes and thus require less computational effort.  

\subsection{Tree Distribution approximation}

TDA employs SDP to approximate the final cascade size distribution on a general network $G = (V, E)$. 
First, we compute a minimum spanning tree $M$ of $G$ and run $SDP$ on $M$ with updated response functions $\tilde{R}_i$. 
The algorithms consists of four main steps that are detailed next.  
1) We first compute the activation probability $p_i$ of each node $i \in V$ by belief propagation on $G$. 
We therefore need to know how many neighbors activate before the node $i$. 
Each neighbor $j$ activates with probability $p_{ij} =  \mathbb{P}\left(s_i = 1 \parallel s_j = 0\right)$ before $i$ and, according to our BP assumption, all neighbors activate independently. 
They fulfill the self-consistent equations:
\begin{align*}
p_{ij} & =  \mathbb{P}\left(s_i = 1 \parallel s_j = 0\right) = \sum_{\mathbf{s}_{ \rm{nb}(i)\setminus {j}} \in \{0,1\}^{d_i-1}} R^c_i\left(\sum_{n \in \rm{nb}(i)\setminus {j}} s_n \right) \prod_{n \in \rm{nb}(i)\setminus {j}} p^{s_n}_{ni} (1- p_{ni})^{1-s_n}, 
\end{align*}
where $\rm{nb}(i)$ denotes the set of neighbors of $i$ and $\mathbf{s}_{ \rm{nb}(i)\setminus {j}}$ a vector consisting of states $s_n$ of $i$'s neighbors $n$ except $j$. 
If $G$ is a tree, the independence assumption is correct and we only need to visit each node twice to calculate the correct probabilities $p_{ij}$. 
Starting in the bottom of a tree, for each node $n$, we can compute $p_{np}$ based on $n$'s children, while its parent $p$ has no influence on $n$.  
Next, we start in the root of the tree and proceed to compute $p_{pn}$ until we reach the bottom. 
However, this is not enough if $G$ is not a tree. 
Then, loopy BP interprets the equation above as system of fixed point equations (for $p_{ij}$) that we solve iteratively.
A reasonable initialization is $p_{ij}  = R_i(0)$. 
For TDA, we always iterate $50$ times through the whole network, which is enough to reach convergence in our cases. 
The product over neighbors is computed efficiently with the help of Fast Fourier Transformations. 
Based on $p_{ij}$, the activation probability of a node reads as 
\begin{align*}
p_{i}  =  \mathbb{P}\left(s_i = 1 \right)  = \sum_{\mathbf{s}_{ \rm{nb}(i)} \in \{0,1\}^{d_i}} R^c_i\left(\sum_{n \in \rm{nb}(i)} s_n \right) \prod_{n \in \rm{nb}(i)} p^{s_n}_{ni} (1- p_{ni})^{1-s_n}. 
\end{align*}
2) We compute a minimum spanning tree $M = (V_M, E_M)$ of the original network $G$. 
We report results for a randomly chosen minimum spanning tree. 
However, weighting edges can give preference to which edges should be removed or kept, for instance, edges connecting nodes with larger degrees etc.
Let us denote by $\rm{dnb}(i) = \{j \in V \mid  (i,j) \in E,  \; (i,j) \notin E_M\}$ the set of neighbors of a node $i$ in $G$ that $i$ is not connected to anymore in $M$, and let $m_i = |\rm{dnb}(i)| $ be the number of such lost neighbors. 
3) Then, we update the response functions $R_i$ of each node $i$ by the probability that $i$ activates after $a$ of its neighbors in $M$ activated. 
In addition, we assume that each of $i$'s deleted neighbors $n$ has activated initially with probability $p_{ni}$.
We therefore consider the activation of the deleted neighbors as independent of the rest of the cascade. 
Accordingly, $R_i(a)$ is defined as average with respect to initial failures of deleted neighbors:   
\begin{align*}
\tilde{R}_i(a) = \sum_{\mathbf{s}_{\rm{dnb}(i)} \in \{0,1\}^{m_i}} R_i\left(a + \sum_{n \in \rm{dnb}(i)} s_n \right) \prod_{n \in \rm{dnb}(i)}  p^{s_n}_{ni} (1- p_{ni})^{1-s_n} . 
\end{align*} 
4) Finally, the cascade size distribution is computed by SDP with inputs $M$ and $\tilde{R}_i$.

\end{methods}


\begin{addendum}
\item[Supplementary Information] is provided alongside the main manuscript.
 \item[Acknowledgments] R.B. acknowledges support by the ETH48 project of the ETH Risk Center and thanks Frank Schweitzer and the Chair of Systems Design at ETH Zurich for their generous hospitality.
 \item[Competing Interests] The author declares that she has no
competing financial interests.
 \item[Correspondence] Correspondence and requests for materials
should be addressed to R.B.\\ (email: rburkholz@ethz.ch).
\end{addendum}


\begin{thebibliography}{28}
\expandafter\ifx\csname url\endcsname\relax
  \def\url#1{\texttt{#1}}\fi
\expandafter\ifx\csname urlprefix\endcsname\relax\def\urlprefix{URL }\fi
\providecommand{\bibinfo}[2]{#2}
\providecommand{\eprint}[2][]{\url{#2}}

\bibitem{Newman.Strogatz.ea2001Randomgraphswith}
\bibinfo{author}{Newman, M. E.~J.}, \bibinfo{author}{Strogatz, S.~H.} \&
  \bibinfo{author}{Watts, D.~J.}
\newblock \bibinfo{title}{{Random graphs with arbitrary degree distributions
  and their applications}}.
\newblock \emph{\bibinfo{journal}{Physcal Review E}}
  \textbf{\bibinfo{volume}{64}}, \bibinfo{pages}{026118}
  (\bibinfo{year}{2001}).

\bibitem{brain2}
\bibinfo{author}{Kinouchi, O.} \& \bibinfo{author}{Copelli, M.}
\newblock \bibinfo{title}{Optimal dynamical range of excitable networks at
  criticality} \textbf{\bibinfo{volume}{2}}, \bibinfo{pages}{348--351}
  (\bibinfo{year}{2006}).
\newblock
  \urlprefix\url{https://app.dimensions.ai/details/publication/pub.1018095608}.

\bibitem{brain}
\bibinfo{author}{Friedman, N.} \emph{et~al.}
\newblock \bibinfo{title}{Universal critical dynamics in high resolution
  neuronal avalanche data}.
\newblock \emph{\bibinfo{journal}{Phys. Rev. Lett.}}
  \textbf{\bibinfo{volume}{108}}, \bibinfo{pages}{208102}
  (\bibinfo{year}{2012}).
\newblock
  \urlprefix\url{https://link.aps.org/doi/10.1103/PhysRevLett.108.208102}.

\bibitem{NewmanSIR}
\bibinfo{author}{Newman, M. E.~J.}
\newblock \bibinfo{title}{Spread of epidemic disease on networks}.
\newblock \emph{\bibinfo{journal}{Physical Review E}}
  \textbf{\bibinfo{volume}{66}}, \bibinfo{pages}{016128}
  (\bibinfo{year}{2002}).
\newblock \urlprefix\url{https://link.aps.org/doi/10.1103/PhysRevE.66.016128}.

\bibitem{Battiston2012a}
\bibinfo{author}{Battiston, S.}, \bibinfo{author}{{Delli Gatti}, D.},
  \bibinfo{author}{Gallegati, M.}, \bibinfo{author}{Greenwald, B. C.~N.} \&
  \bibinfo{author}{Stiglitz, J.~E.}
\newblock \bibinfo{title}{{Credit Default Cascades: When Does Risk
  Diversification Increase Stability?}}
\newblock \emph{\bibinfo{journal}{Journal of Financial Stability}}
  \textbf{\bibinfo{volume}{8}}, \bibinfo{pages}{138--149}
  (\bibinfo{year}{2012}).

\bibitem{Watts2002}
\bibinfo{author}{Watts, D.~J.}
\newblock \bibinfo{title}{{A simple model of global cascades on random
  networks}}.
\newblock \emph{\bibinfo{journal}{Proceedings of the National Academy of
  Sciences USA}} \textbf{\bibinfo{volume}{99}}, \bibinfo{pages}{5766--5771}
  (\bibinfo{year}{2002}).

\bibitem{Burkholz2015}
\bibinfo{author}{Burkholz, R.}, \bibinfo{author}{Garas, A.} \&
  \bibinfo{author}{Schweitzer, F.}
\newblock \bibinfo{title}{How damage diversification can reduce systemic risk}.
\newblock \emph{\bibinfo{journal}{Physical Review E}}
  \textbf{\bibinfo{volume}{93}}, \bibinfo{pages}{042313}
  (\bibinfo{year}{2016}).
\newblock \urlprefix\url{http://link.aps.org/doi/10.1103/PhysRevE.93.042313}.

\bibitem{BurkholzMultiplex}
\bibinfo{author}{Burkholz, R.}, \bibinfo{author}{Leduc, M.~V.},
  \bibinfo{author}{Garas, A.} \& \bibinfo{author}{Schweitzer, F.}
\newblock \bibinfo{title}{Systemic risk in multiplex networks with asymmetric
  coupling and threshold feedback}.
\newblock \emph{\bibinfo{journal}{Physica D: Nonlinear Phenomena}}
  \textbf{\bibinfo{volume}{323–324}}, \bibinfo{pages}{64 -- 72}
  (\bibinfo{year}{2016}).
\newblock
  \urlprefix\url{http://www.sciencedirect.com/science/article/pii/S0167278915001943}.
\newblock \bibinfo{note}{Nonlinear Dynamics on Interconnected Networks}.

\bibitem{LRD}
\bibinfo{author}{Burkholz, R.} \& \bibinfo{author}{Schweitzer, F.}
\newblock \bibinfo{title}{Framework for cascade size calculations on random
  networks}.
\newblock \emph{\bibinfo{journal}{Physical Review E}}
  \textbf{\bibinfo{volume}{97}}, \bibinfo{pages}{042312}
  (\bibinfo{year}{2018}).
\newblock \urlprefix\url{https://link.aps.org/doi/10.1103/PhysRevE.97.042312}.

\bibitem{BurkholzCorrelations}
\bibinfo{author}{Burkholz, R.} \& \bibinfo{author}{Schweitzer, F.}
\newblock \bibinfo{title}{Correlations between thresholds and degrees: An
  analytic approach to model attacks and failure cascades}.
\newblock \emph{\bibinfo{journal}{Physical Review E}}
  \textbf{\bibinfo{volume}{98}}, \bibinfo{pages}{022306}
  (\bibinfo{year}{2018}).
\newblock \urlprefix\url{https://link.aps.org/doi/10.1103/PhysRevE.98.022306}.

\bibitem{BianconiExtreme}
\bibinfo{author}{Bianconi, G.}
\newblock \bibinfo{title}{Rare events and discontinuous percolation
  transitions}.
\newblock \emph{\bibinfo{journal}{Physical Review E}}
  \textbf{\bibinfo{volume}{97}}, \bibinfo{pages}{022314}
  (\bibinfo{year}{2018}).
\newblock \urlprefix\url{https://link.aps.org/doi/10.1103/PhysRevE.97.022314}.

\bibitem{BP}
\bibinfo{author}{Braunstein, A.}, \bibinfo{author}{Mézard, M.} \&
  \bibinfo{author}{Zecchina, R.}
\newblock \bibinfo{title}{Survey propagation: An algorithm for satisfiability}.
\newblock \emph{\bibinfo{journal}{Random Structures \& Algorithms}}
  \textbf{\bibinfo{volume}{27}}, \bibinfo{pages}{201--226}.
\newblock
  \urlprefix\url{https://onlinelibrary.wiley.com/doi/abs/10.1002/rsa.20057}.

\bibitem{BurkholzFinite}
\bibinfo{author}{Burkholz, R.}, \bibinfo{author}{Herrmann, H.~J.} \&
  \bibinfo{author}{Schweitzer, F.}
\newblock \bibinfo{title}{Explicit size distributions of failure cascades
  redefine systemic risk on finite networks}.
\newblock \emph{\bibinfo{journal}{Scientific Reports}} \bibinfo{pages}{1--8}
  (\bibinfo{year}{2018}).
\newblock \urlprefix\url{https://rdcu.be/NbZe}.

\bibitem{friedli_velenik_2017}
\bibinfo{author}{Friedli, S.} \& \bibinfo{author}{Velenik, Y.}
\newblock \emph{\bibinfo{title}{Statistical Mechanics of Lattice Systems: A
  Concrete Mathematical Introduction}} (\bibinfo{publisher}{Cambridge
  University Press}, \bibinfo{year}{2017}).

\bibitem{ThurnerFinance}
\bibinfo{author}{Poledna, S.} \& \bibinfo{author}{Thurner, S.}
\newblock \bibinfo{title}{Elimination of systemic risk in financial networks by
  means of a systemic risk transaction tax}.
\newblock \emph{\bibinfo{journal}{Quantitative Finance}}
  \textbf{\bibinfo{volume}{16}}, \bibinfo{pages}{1599--1613}
  (\bibinfo{year}{2016}).
\newblock \urlprefix\url{https://doi.org/10.1080/14697688.2016.1156146}.

\bibitem{BPthresholds}
\bibinfo{author}{Gleeson, J.~P.} \& \bibinfo{author}{Porter, M.~A.}
\newblock \emph{\bibinfo{title}{Complex Spreading Phenomena in Social Systems}}
  (\bibinfo{publisher}{Springer}, \bibinfo{year}{2018}).

\bibitem{GleesonPairApproximation}
\bibinfo{author}{Gleeson, J.~P.}
\newblock \bibinfo{title}{Binary-state dynamics on complex networks: Pair
  approximation and beyond}.
\newblock \emph{\bibinfo{journal}{Physical Review X}}
  \textbf{\bibinfo{volume}{3}}, \bibinfo{pages}{021004} (\bibinfo{year}{2013}).
\newblock \urlprefix\url{https://link.aps.org/doi/10.1103/PhysRevX.3.021004}.

\bibitem{wattsCollect}
\bibinfo{author}{Watts, D.~J.} \& \bibinfo{author}{Dodds, P.~S.}
\newblock \bibinfo{title}{Threshold models of social influence}.
\newblock In \emph{\bibinfo{booktitle}{The Oxford Handbook of Analytical
  Sociology}}, chap.~\bibinfo{chapter}{20}, \bibinfo{pages}{475--497}
  (\bibinfo{publisher}{Oxford University Press}, \bibinfo{address}{Oxford, UK},
  \bibinfo{year}{2009}).

\bibitem{Granovetter1978}
\bibinfo{author}{Granovetter, M.~S.}
\newblock \bibinfo{title}{{Threshold Models of Collective Behavior}}.
\newblock \emph{\bibinfo{journal}{The American Journal of Sociology}}
  \textbf{\bibinfo{volume}{83}}, \bibinfo{pages}{1420--1443}
  (\bibinfo{year}{1978}).

\bibitem{Kermack700}
\bibinfo{author}{Kermack, W.~O.} \& \bibinfo{author}{McKendrick, A.~G.}
\newblock \bibinfo{title}{A contribution to the mathematical theory of
  epidemics}.
\newblock \emph{\bibinfo{journal}{Proceedings of the Royal Society of London A:
  Mathematical, Physical and Engineering Sciences}}
  \textbf{\bibinfo{volume}{115}}, \bibinfo{pages}{700--721}
  (\bibinfo{year}{1927}).
\newblock
  \urlprefix\url{http://rspa.royalsocietypublishing.org/content/115/772/700}.

\bibitem{Kempe2003}
\bibinfo{author}{Kempe, D.}, \bibinfo{author}{Kleinberg, J.} \&
  \bibinfo{author}{Tardos, {\'{E}}.}
\newblock \bibinfo{title}{{Maximizing the Spread of Influence Through a Social
  Network}}.
\newblock In \emph{\bibinfo{booktitle}{Proceedings of the Ninth ACM SIGKDD
  International Conference on Knowledge Discovery and Data Mining}}, KDD '03,
  \bibinfo{pages}{137--146} (\bibinfo{publisher}{ACM}, \bibinfo{address}{New
  York, NY, USA}, \bibinfo{year}{2003}).
\newblock \urlprefix\url{http://doi.acm.org/10.1145/956750.956769}.

\bibitem{Jordan}
\bibinfo{editor}{Jordan, M.~I.} (ed.) \emph{\bibinfo{title}{Learning in
  Graphical Models}} (\bibinfo{publisher}{MIT Press},
  \bibinfo{address}{Cambridge, MA, USA}, \bibinfo{year}{1999}).

\bibitem{EVAData}
\bibinfo{author}{Norlen, K.}, \bibinfo{author}{Lucas, G.},
  \bibinfo{author}{Gebbie, M.} \& \bibinfo{author}{Chuang, J.}
\newblock \bibinfo{title}{{EVA: Extraction, Visualization and Analysis of the
  Telecommunications and Media Ownership Network}} (\bibinfo{year}{2002}).

\bibitem{Gleeson2007}
\bibinfo{author}{Gleeson, J.~P.} \& \bibinfo{author}{Cahalane, D.}
\newblock \bibinfo{title}{{Seed size strongly affects cascades on random
  networks}}.
\newblock \emph{\bibinfo{journal}{Physical Review E}}
  \textbf{\bibinfo{volume}{75}}, \bibinfo{pages}{1--4} (\bibinfo{year}{2007}).

\bibitem{PhysRevX.4.021024}
\bibinfo{author}{Altarelli, F.}, \bibinfo{author}{Braunstein, A.},
  \bibinfo{author}{Dall'Asta, L.}, \bibinfo{author}{Wakeling, J.~R.} \&
  \bibinfo{author}{Zecchina, R.}
\newblock \bibinfo{title}{Containing epidemic outbreaks by message-passing
  techniques}.
\newblock \emph{\bibinfo{journal}{Physical Review X}}
  \textbf{\bibinfo{volume}{4}}, \bibinfo{pages}{021024} (\bibinfo{year}{2014}).
\newblock \urlprefix\url{https://link.aps.org/doi/10.1103/PhysRevX.4.021024}.

\bibitem{DBLP:conf/nips/Lokhov16}
\bibinfo{author}{Lokhov, A.~Y.}
\newblock \bibinfo{title}{Reconstructing parameters of spreading models from
  partial observations}.
\newblock In \emph{\bibinfo{booktitle}{Advances in Neural Information
  Processing Systems 29, Dec 5-10, Barcelona, Spain}},
  \bibinfo{pages}{3459--3467} (\bibinfo{year}{2016}).
\newblock
  \urlprefix\url{http://papers.nips.cc/paper/6129-reconstructing-parameters-of-spreading-models-from-partial-observations}.

\bibitem{Barabasi1999}
\bibinfo{author}{Barab{\'{a}}si, A.~L.} \& \bibinfo{author}{Albert, R.}
\newblock \bibinfo{title}{{Emergence of Scaling in Random Networks}}.
\newblock \emph{\bibinfo{journal}{Science}} \textbf{\bibinfo{volume}{286}},
  \bibinfo{pages}{509--512} (\bibinfo{year}{1999}).
\newblock \urlprefix\url{http://www.sciencemag.org/content/286/5439/509}.

\bibitem{Clauset2009}
\bibinfo{author}{Clauset, A.}, \bibinfo{author}{Shalizi, C.~R.} \&
  \bibinfo{author}{Newman, M. E.~J.}
\newblock \bibinfo{title}{{Power-Law Distributions in Empirical Data}}.
\newblock \emph{\bibinfo{journal}{Society for Industrial and Applied
  Mathematics Review}} \textbf{\bibinfo{volume}{51}}, \bibinfo{pages}{661}
  (\bibinfo{year}{2009}).
\newblock
  \urlprefix\url{http://link.aip.org/link/SIREAD/v51/i4/p661/s1{\&}Agg=doi}.

\end{thebibliography}
\end{document}